\documentclass[a4paper,11pt]{article}
\pdfoutput=1 % if your are submitting a pdflatex (i.e. if you have
\usepackage{jheppub} % for details on the use of the package, please
\usepackage[T1]{fontenc} % if needed

\usepackage{MnSymbol}

\usepackage{amsmath}

	\newcommand\SmallMatrix[1]{{%
  {\tiny{\arraycolsep=0.3\arraycolsep\ensuremath{\left\{\begin{matrix}#1\end{matrix}\right\}}}}}}
	
\title{\boldmath Leading multi-soft limits from scattering equations}

\author[a]{Michael Zlotnikov}

\affiliation[a]{Brown University\\Department of Physics\\182 Hope St, Providence, RI, 02912}

\emailAdd{michael\_zlotnikov@brown.edu}

\abstract{\\A Cachazo-He-Yuan (CHY) type formula is derived for the leading gluon, bi-adjoint scalar $\phi^3$, Yang-Mills-scalar and non-linear sigma model $m$-soft factors $S_m$ in arbitrary dimension. The general formula is used to evaluate explicit examples for up to three soft legs analytically and up to four soft legs numerically via comparison with amplitude ratios under soft kinematics. A structural pattern for gluon $m$-soft factor is inferred and a simpler formula for its calculation is conjectured. In four dimensions, a Cachazo-Svr\v{c}ek-Witten (CSW) recursive procedure producing the leading $m$-soft gluon factor in spinor helicity formalism is developed as an alternative, and Britto-Cachazo-Feng-Witten (BCFW) recursion is used to obtain the leading four-soft gluon factor for all analytically distinct helicity configurations.}

\begin{document} 
\maketitle
\flushbottom

\section{Introduction}
		Investigation of soft factors has a rich history, reaching back to the contributions of Low, Weinberg and others \cite{LowFourPt1,LowFourPt2,LowFourPt3,LowTheorem,Weinberg1,weinberg,OtherSoftPhotons1,Gross:1968in,OtherSoftPhotons2,OtherSoftPhotons3}. Soft factorization is a universal property of scattering amplitudes. An $n$-point scattering amplitude $A_n$ depends on external momenta $k_i^\mu$ of the $i=1,2,...,n$ ingoing and outgoing scattering particles. If a subset of adjacent external momenta $k_j^\mu$ for $\forall j=1,2,...,m$ with $m<n-3$ is taken to zero, for example parametrized as $k_j^\mu\to\tau k_j^\mu$ and $\tau\to 0$, the amplitude is expected to factorize at leading order in $\tau$ into a soft factor $S_m$ times a lower point amplitude $A_{n-m}$:
		\begin{align}
A_n\to S_m A_{n-m}+\text{sub-leading in }\tau.
\end{align}
		Universality in this context means that $S_m$ is independent of the remaining lower point amplitude $A_{n-m}$, such that $S_m$ is always the same whenever the same types of $m$ external particles are taken soft within any original amplitude $A_n$.
		
		More recently, interest in investigation of soft theorems was refueled \cite{strominger1,strominger2,strominger3,Cachazo:2014fwa} as Strominger et al. showed that soft-graviton theorems can be understood from the point of view of BMS symmetry \cite{BMS1,BMS2,ExtendedBMS1,ExtendedBMS2,ExtendedBMS3}. Further study of leading and sub-leading soft theorems in Yang-Mills, gravity, string and supersymmetric theories ensued \cite{Cachazo:2013hca,schwab,nima,1405.1015,1405.1410,Broedel:2014fsa,Bern:2014vva,Casali:2014xpa,Larkoski:2014hta,1406.7184,Zlotnikov:2014sva,Kalousios:2014uva,1408.4179,Lipstein:2015rxa,Avery:2015gxa,1509.07840,Bianchi:2015lnw,Avery:2015iix,Bianchi:2016tju,DiVecchia:2016amo,1509.01406,1605.09094,1608.00685,1610.03481,1702.03934,1703.00024,1706.00759,Chakrabarti:2017ltl,Mao:2017wvx,1611.02172,1611.07534}, partly based on the amplitude formulation due to Cachazo, He and Yuan (CHY) \cite{Cachazo:2013hca}. Double soft theorems have been considered in \cite{weinberg2,JS,adler}, and more recently \cite{ArkaniHamed:2008gz,Chen:2014cuc,Chen:2014xoa,Cachazo:2015ksa,Volovich:2015yoa,Klose:2015xoa,Du:2015esa,DiVecchia:2015bfa,1507.08829,1505.05854,1502.05258,1406.4172,1406.5155,1411.6661,DiVecchia:2015srk,Low:2015ogb,He:2016vfi,Saha:2016kjr,Saha:2017yqi,1604.02834,1705.06175}. Construction rules for soft factors with multiple soft particles in $\mathcal{N}=4$ SYM theory appeared in \cite{Georgiou:2015jfa}. Work on related topics was also done, like sub-leading collinear limits \cite{Nandan:2016ohb} and investigation of the current algebra at null infinity induced by soft gluon limits \cite{McLoughlin:2016uwa}.
		
In this note we use the CHY formulation of scattering amplitudes \cite{Cachazo:2013hca,Cachazo:2014xea} to derive the leading $m$-soft factor $S_m$ for gluons, bi-adjoint scalar $\phi^3$, Yang-Mills-scalar and non-linear sigma model.

We find the $m$-soft gluon factor in the case when external legs $1,2,...,m$ are soft to be given by the CHY type formula (\ref{softfacres}, \ref{softmeasure}, \ref{softscat}, \ref{softps}). We then consider explicit examples, obtain analytic results in cases $m=1,2,3\,$, and check the cases $m=2,3,4$ numerically via amplitude ratios in four dimensions obtained from the GGT package \cite{Dixon:2010ik}. Based on these explicit examples, we infer and conjecture a general pattern for the $m$-soft gluon factor:
\begin{align}
\label{smres0}
S^{gluon}_m&=\sum_{r=1}^{m+1}(-1)^{r+1}P^{(m+1-r)}_{r,r+1,...,m,m+1}P^{(r-1)}_{r-1,r-2,...,1,n}\,,
\end{align}
where $P^{(0)}_{m+1}=P^{(0)}_{n}\equiv 1$, and $P^{(i)}_{1,2,...,i,i+1}$, with $d\nu_1$ and $\psi_{[1,i]}^{(i+1)}$ defined in (\ref{softmeasure}) and (\ref{softps}), is\footnote{The cases $P^{(m+1-r)}_{r,r+1,...,m,m+1}$ and $P^{(i)}_{i,i-1,...,2,1,n}$ are obtained by simple index exchange after integration.}
\begin{align}
\label{Pdef0}
P^{(i)}_{1,2,...,i,i+1}=&\int d\nu_1\frac{1}{\prod_{c=2}^{i+1}\bar\sigma_{c-1,c}}\text{Pf}\left(\psi_{[1,i]}^{(i+1)}\right).
\end{align}
If all $P^{(i)}_{1,2,...,i,i+1}$ with $i<m$ are known from calculations of lower soft factors, then $P^{(m)}_{1,2,...,m,m+1}$ is the only new contribution that has to be computed to construct $S_m$ at a given $m$.

 The leading $m$-soft factor in bi-adjoint scalar $\phi^3$, Yang-Mills-scalar and non-linear sigma model theories involves the same integration measure $d\nu_r$ as in (\ref{softfacres}), while the integrands are different: (\ref{SmScalar}), (\ref{SmYMS}) and (\ref{SmNLSM}).

As an alternative in four dimensions, we also develop a CSW type \cite{Cachazo:2004kj} automated recursive procedure that gives the leading $m$-soft gluon factor (compare with construction rules in \cite{Georgiou:2015jfa}). Finally, we use BCFW recursion \cite{Britto:2004ap} to obtain all leading four-soft gluon factors with analytically distinct helicity combinations in four dimensions.

This work is organized as follows. In section \ref{CHYsoft} we recall the CHY formalism and introduce the soft limit. In section \ref{factorization} we demonstrate the soft factorization of gluons at any $m$ and obtain our general result. Explicit examples are worked out in section \ref{examples} and a simpler evaluation formula is conjectured. Multi-soft factors in scalar $\phi^3$, Yang-Mills-scalar and non-linear sigma model are discussed in section \ref{othertheories}. Appendix \ref{CSWrecur} contains a CSW type recursive procedure for $m$-soft factors in four dimensions. Appendix \ref{foursoftBCFW} contains BCFW results for four-soft gluon factors in four dimensions.

\section{The CHY formulation of Yang-Mills and the soft limit}
\label{CHYsoft}
\noindent We start with the usual $n$-point formula for the tree-level gluon amplitude \cite{Cachazo:2013hca}:
\begin{align}
\label{ampl}
A_n=&\int d\mu_n ~{\cal I}_n^{YM},
\end{align}
where the CHY integration measure $d\mu_n$ and the Yang-Mills CHY integrand ${\cal I}_n^{YM}$ are
\begin{align}
\label{CHYmeasure}
d\mu_n=&\int d^n\sigma\frac{\sigma_{ij}\sigma_{jk}\sigma_{ki}}{\text{vol}\left(SL(2,C)\right)}\prod_{\substack{a=1\\a\neq i,j,k}}^n\delta\left(\sum_{\substack{b=1\\b\neq a}}^n\frac{k_a\cdot k_b}{\sigma_{ab}}\right)~~~,~~~{\cal I}_n^{YM}=\frac{2\frac{(-1)^{p+q}}{\sigma_{pq}}\text{Pf}(\Psi^{pq}_{pq})}{\sigma_{12}\sigma_{23}...\sigma_{n1}}.
\end{align}
Moduli differences are abbreviated as $\sigma_{ab}\equiv \sigma_a-\sigma_b$ and the matrix $\Psi$ is given by
\begin{align}
\label{psimat}
\Psi=\left({{A}\atop{C}}~{{-C^T}\atop{B}}\right),~~A=\left\{{{\frac{k_a\cdot k_b}{\sigma_{ab}}}\atop{0}}{{;}\atop{;}}{{a\neq b}\atop{a=b}}\right.,~~B=\left\{{{\frac{\epsilon_a\cdot \epsilon_b}{\sigma_{ab}}}\atop{0}}{{;}\atop{;}}{{a\neq b}\atop{a=b}}\right.,~~C=\left\{\begin{matrix}
\frac{\epsilon_a\cdot k_b}{\sigma_{ab}}&;\, a\neq b\\
-\sum_{\substack{c=1\\c\neq a}}^n\frac{\epsilon_a\cdot k_c}{\sigma_{ac}}&;\, a=b
\end{matrix}\right.,
\end{align}
with $a,b\in\{1,2,...,n\}$. The $k^\mu$ are momenta of scattering particles and $\epsilon^\mu$ contain the corresponding polarization data. The indices $1\leq i<j<k\leq n$ as well as $1\leq p<q\leq n$ in (\ref{CHYmeasure}) are chosen arbitrarily but fixed. Upper and lower indices on matrix $\Psi$ denote removed columns and  rows respectively. We would like to consider the case where $m$ external legs with $m<n-3$ are going soft simultaneously:
\begin{align}
\label{taulim}
k^\mu_q\to \tau k^\mu_q,~~~\tau\to 0,~~~\text{for}~~~q\in \{1,2,...,m\}.
\end{align}
As we take $\tau\to 0$, it is clear from the structure of matrix $\Psi$ that at leading order in $\tau$ the Pfaffian factorizes as:\footnote{To see this, make the substitution (\ref{taulim}) and expand the Pfaffian along rows and/or columns $1,2,...,m,n+1,n+2,...,n+m$. Retain only leading summands under $\tau\to0$, keeping in mind that solutions with $\sigma_{ab}=O(\tau)$ or $\sigma_{ab}=O(1)$ for $a,b\in\{1,2,...,m\}$ are possible. Finally, reassemble the remaining coefficients into Pf$(\psi)$.}
\begin{align}
\label{pffactor}
\text{Pf}(\Psi^{pq}_{pq})\to\text{Pf}(\psi)\text{Pf}(\Psi^{p,q,1,2,...,m,n+1,n+2,...,n+m}_{p,q,1,2,...,m,n+1,n+2,...,n+m}|_{\tau=0})+\text{subleading in }\tau,
\end{align}
possibly up to an overall sign. The $2m\times 2m$ matrix $\psi$ in the first Pfaffian on the right hand side of (\ref{pffactor}) is defined the same way as $\Psi$, except the indices $a,b$ in the sub-matrices $A,B,C$ are restricted to the subset $a,b\in\{1,2,...,m\}$. Here, to do the expansion along rows we employed the usual recursive formula for the Pfaffian of an anti-symmetric $2n\times 2n$ matrix $M$:
\begin{align}
{\rm Pf}\left(M\right)=\sum_{{j=1}\atop{j\neq i}}^{2n}(-1)^{i+j+1+\theta_{i-j}}m_{ij}{\rm Pf}\left(M^{ij}_{ij}\right),
\end{align}
where $m_{ij}$ are elements of matrix $M$, $\theta_x\equiv\theta(x)$ is the Heaviside step function, and index $i$ can be freely chosen.\\
\indent Alternatively, we could have noticed that $\tau\to 0$ reduces matrix $\Psi^{pq}_{pq}$ at leading order to a block matrix structure, with several blocks equal to zero. Factorization (\ref{pffactor}) then directly follows from trivial Pfaffian factorization identities for block matrices.\\
\indent Note that $\text{Pf}(\psi)$ contains terms leading and/or sub-leading in $\tau$, depending on whether it is evaluated on degenerate ($\sigma_{ab}=O(\tau)$ for some $a,b$) or non-degenerate ($\sigma_{ab}=O(1)$ for all $a,b$) solutions to the scattering equations. However, for our purposes it is only important that for all types of solutions $\text{Pf}(\psi)$ contains all leading contributions.\\
\indent The second Pfaffian on the right hand side of (\ref{pffactor}) is the one we expect in an $(n-m)$-point amplitude as we take $\tau\to0$. Furthermore, we can trivially rewrite
\begin{align}
\frac{1}{\sigma_{12}\sigma_{23}...\sigma_{n1}}=\frac{\sigma_{n,m+1}}{\sigma_{n1}\sigma_{12}...\sigma_{m,m+1}}\cdot\frac{1}{\sigma_{n,m+1}\sigma_{m+1,m+2}...\sigma_{n-1,n}},
\end{align}
and observe the following behavior in scattering equation delta functions
\begin{align}
\label{deltafact1}
\prod_{a=1}^n\delta\left(\sum_{{b=1}\atop{b\neq a}}^n\frac{k_a\cdot k_b}{\sigma_{ab}}\right)=\prod_{a=1}^m\delta\left(\sum_{{b=1}\atop{b\neq a}}^n\frac{k_a\cdot k_b}{\sigma_{ab}}\right)\prod_{c=m+1}^n\delta\left(\sum_{{b=m+1}\atop{b\neq c}}^n\frac{k_c\cdot k_b}{\sigma_{cb}}+O(\tau)\right).
\end{align}
The last equation holds since we necessarily have $\sigma_{cb}=O(1)$ for $ m+1\leq c\leq n$ due to the kinematics in all $k^\mu_c$ being generic and therefore producing non-degenerate configurations of $\sigma_c$, while all $k_b=O(\tau)$ for the soft particles $1\leq b\leq m$ tend to zero. The behavior of the first $1\leq a\leq m$ delta functions in (\ref{deltafact1}) is more subtle, since we can have $\sigma_{ab}=O(1)$ or $\sigma_{ab}=O(\tau)$ in this case. It will be investigated in detail in the next section.\\
\indent Considering the above, we can structurally rewrite (\ref{ampl}) at leading order in $\tau\to 0$ as
\begin{align}
A_n\to&\int d\mu_{n-m} ~S_m~{\cal I}_{n-m}^{YM}+\text{sub-leading in }\tau,\\
\label{sdef}
S_m=&\int d^m\sigma\prod_{a=1}^m\delta\left(\sum_{{b=1}\atop{b\neq a}}^n\frac{k_a\cdot k_b}{\sigma_{ab}}\right)\frac{\sigma_{n,m+1}}{\sigma_{n1}\sigma_{12}...\sigma_{m,m+1}}\text{Pf}(\psi),
\end{align}
where $d\mu_{n-m}$ and ${\cal I}_{n-m}^{YM}$ are based on objects with indices in the range $\{m+1,m+2,...,n\}$. \\
\indent Of course this alone does not provide a factorization yet, since $S_m$ still depends on $\sigma_n$ and $\sigma_{m+1}$, and the delta functions within still depend on all $n$ momenta and $\sigma$-moduli. In the following we show that for any $m$ the $\sigma_{m+1},...,\sigma_n$ dependence in $S_m$ drops out at leading order in $\tau$ and the amplitude indeed factorizes as $A_n\to S_m A_{n-m}+$ sub-leading in $\tau$. Furthermore, we find that $S_m$ only depends on polarizations $\epsilon_1^\mu,\epsilon_2^\mu,...,\epsilon_m^\mu$ as well as momenta $k_n^\mu,k_1^\mu,k_2^\mu,...,k_{m+1}^\mu$, and establish a CHY type formula for evaluating $S_m$ independently of the remaining factored amplitude $A_{n-m}$.

\section{Factorization of $S_m$ for Yang-Mills and the general result}
\label{factorization}
Starting with $S_m$ in (\ref{sdef}) we apply several transformations in order to more conveniently work with this expression. First we rewrite the delta functions making use of the general identity
\begin{align}
\label{deltatrafo}
\int d^m x\prod_{i=1}^m\delta\left(f_i(\vec x)\right)\,\bullet~ =\int d^m x\det(M)\prod_{i=1}^m\delta\left(\sum_{j=1}^m M_{ij}f_j(\vec x)\right)\,\bullet~,
\end{align}
where $\bullet$ is a placeholder for some test function and we employ the specific $m\times m$ matrix
\begin{align}
M=\left(\begin{matrix}
1 & 1 & 1 & ... & 1 & 1 & 1 \\
1 & -1 & 0 & ... & 0 & 0 & 0 \\
0 & 1 & -1 & ... & 0 & 0 & 0 \\
\ddots & \ddots & \ddots & \ddots & \ddots & \ddots & \ddots \\
0 &0 & 0 & ... & 1 & -1 & 0 \\
0 & 0 & 0 & ... & 0 & 1 & -1 
\end{matrix}\right),~~~~~~\det(M)=(-1)^{m+1}m,
\end{align}
which for our particular variables and functions of interest yields the effective relation
\begin{align}
\label{deltraf}
\prod_{a=1}^m\delta\left(\sum_{{b=1}\atop{b\neq a}}^n\frac{k_a\cdot k_b}{\sigma_{ab}}\right)=&(-1)^{m+1}m\,\delta\left(\sum_{a=1}^m\sum_{b=m+1}^n\frac{k_a\cdot k_b}{\sigma_{ab}}\right)\prod_{q=1}^{m-1}\delta\left(h_q\right),\\
\label{hdef}
h_q=& \sum_{{a=1}\atop{a\neq q}}^n\frac{k_q\cdot k_a}{\sigma_{qa}}-\sum_{{b=1}\atop{b\neq q+1}}^n\frac{k_{q+1}\cdot k_b}{\sigma_{q+1,b}}.
\end{align}
Furthermore, we transform the moduli $\sigma_a$ into a new set of variables $\rho$ and $\xi_i$:
\begin{align}
\label{sigtraf}
\sigma_q=\rho-\sum_{a=1}^{q-1}\frac{\xi_a}{2}+\sum_{b=q}^{m-1}\frac{\xi_b}{2},
\end{align}
which leads to a change of the integration measure as
\begin{align}
d\sigma_1\wedge d\sigma_2\wedge ...\wedge d\sigma_m =(-1)^{m+1}d\rho\wedge d\xi_1\wedge d\xi_2\wedge...\wedge d\xi_{m-1}.
\end{align}
The transformation (\ref{sigtraf}) is convenient, since $\sigma_{a,a+1}=\xi_a$ allows for more direct access to degenerate solutions $\sigma_{a,a+1}=O(\tau)$ in the new $\xi_a$ variables. To keep expressions short, we will maintain the $\sigma_a$ notation while implying the substitution (\ref{sigtraf}). With the above changes, $S_m$ becomes
\begin{align}
S_m=\int d\rho\, d^{m-1}\xi ~m\,\delta\left(\sum_{a=1}^m\sum_{b=m+1}^n\frac{k_a\cdot k_b}{\sigma_{ab}}\right)\prod_{q=1}^{m-1}\delta\left(h_q\right)\frac{\sigma_{n,m+1}}{\sigma_{n1}\sigma_{12}...\sigma_{m,m+1}}\text{Pf}(\psi).
\end{align}

Now consider keeping $\rho$ fixed and integrating out the $q=1,2,...,m-1$ constraints $h_q=0$ (which we will denote as $\{h\}=0$) in the $\xi$ variables. This introduces a Jacobian $\det(H)^{-1}$ with derivative matrix $H_{ij}=\partial_{\xi_i}h_j$ and a summation over all solutions to the set of $m-1$ equations $\{h\}=0$ in the $\xi$ variables:
\begin{align}
S_m=\sum_{{\{h\}=0}\atop{\text{solutions}}}\int d\rho \frac{m}{\det(H)}\,\delta\left(\sum_{a=1}^m\sum_{b=m+1}^n\frac{k_a\cdot k_b}{\sigma_{ab}}\right)\frac{\sigma_{n,m+1}}{\sigma_{n1}\sigma_{12}...\sigma_{m,m+1}}\text{Pf}(\psi).
\end{align}
Clearly, here all expressions in the integrand can be effectively thought of as functions of the single variable $\rho$, since $\sigma_a=\sigma_a\left(\rho,\{\xi(\rho)\}\right)$ for $a\in\{1,2,...,m\}$ for each solution of $\{h\}=0$ in $\xi$ variables. Therefore, we can now map the single remaining delta function to a simple pole
\begin{align}
\label{scontour}
S_m=\sum_{{\{h\}=0}\atop{\text{solutions}}}\oint \frac{d\rho}{2\pi i} \frac{m}{\det(H)}\frac{1}{\sum_{a=1}^m\sum_{b=m+1}^n\frac{k_a\cdot k_b}{\sigma_{ab}}}\frac{\sigma_{n,m+1}}{\sigma_{n1}\sigma_{12}...\sigma_{m,m+1}}\text{Pf}(\psi),
\end{align}
 and consider contour deformations away from the initial locus $\sum_{a=1}^m\sum_{b=m+1}^n\frac{k_a\cdot k_b}{\sigma_{ab}}=0$ in $\rho$.

By simple power counting of poles we see that there is no pole and therefore no residue at infinity in $\rho$. As we deform the contour in $\rho$, the expressions $\{h\}$ change dynamically since they depend on $\rho$ directly and through $\xi(\rho)$ variables. When we localize $\rho$ at a pole contained in the integrand, the $\{h\}=0$ constraints can get rescaled and simplified. However, since we are summing over the solutions, the set of constraints $\{h\}=0$ has to stay analytic to leading order at the poles in $\rho$ at all times. This implies i.e. that the Jacobian $\det(H)^{-1}$ can get rescaled and simplified due to the contour deformation, but may never diverge. This is a powerful constraint that allows us to find all integrand poles in $\rho$ as follows.

Structurally, the only type of poles that exists in the integrand is of the shape $1/\sigma_{ab}$. As one such pole becomes localized, corresponding terms in the set of expressions $\{h\}$ start to diverge. Maintaining analyticity at leading order of the divergence in one of the $\{h\}=0$ constraints then demands that at least one different independent $1/\sigma_{cd}$ pole must become localized as well simultaneously and at the same rate.\footnote{This is the case since $\frac{k_a\cdot k_b}{\bar\sigma_{ab}}=0$ for generic momenta only has the solution $|\bar\sigma_{ab}|=\infty$, which is non-analytic, while in the case of at least two summands $\frac{k_a\cdot k_b}{\bar\sigma_{ab}}+\frac{k_c\cdot k_d}{\bar\sigma_{cd}}=0$ finite solutions for the $\bar\sigma_i$ exist such that analyticity is preserved.} This second pole then threatens the analyticity in another $\{h\}=0$ constraint which is affected only by this new divergence, etc. In this fashion a chain of relations occurs demanding that more and more poles must be localized at the same rate simultaneously until it is ensured that analyticity in all $\{h\}=0$ constraints at leading order in the poles is preserved. Overall we realize that whenever a $1/\sigma_{ab}$ pole is localized due to the d.o.f. in $\rho$, the $\rho$ dependence in contributing $\{\xi(\rho)\}$ solutions must be such that other $(m-1)$ independent poles become localized as well simultaneously to maintain analyticity in all the $\{h\}=0$ constraints at leading order of divergence.

Equipped with the above observations, we must consider simultaneously localizing subsets of $m$ independent $1/\sigma_{ab}$ poles in the integrand, with $a\neq b$ pairs $a,b\in\{n,1,2,...,m,m+1\}$. The only part of the integrand which can diverge more or less dependent on the particular choice of $m$ localized independent $1/\sigma_{ab}$ poles is the Parke-Taylor like factor, while all other terms have a fixed scaling (for a given integer $m$). In the following we consider the case of highest divergence, where combinations of $m$ poles that are present in the Parke-Taylor like factor are localized. There are  $\binom{m+1}{m}=m+1$ such pole combinations. We will see that this leads to a simple pole overall, such that any other combination of $m$ localized poles does not develop an overall divergence or residue and thus does not contribute. Therefore, localizing $m$-pole combinations that are present in the Parke-Taylor like factor gives the only non-vanishing contributions.

In the $\rho$ and $\xi_i$ variables the Parke-Taylor like factor reads:
\begin{align}
\label{ptlike}
\frac{\sigma_{n,m+1}}{\sigma_{n1}\sigma_{12}...\sigma_{m,m+1}}=\frac{\sigma_{n,m+1}}{\left(\sigma_{n}-\rho-\sum_{i=1}^{m-1}\frac{\xi_i}{2}\right)\xi_1\xi_2...\xi_{m-1}\left(\rho-\sum_{i=1}^{m-1}\frac{\xi_i}{2}-\sigma_{m+1}\right)}.
\end{align}
Structurally, there are three different classes of $m$-poles combinations that can occur, namely where all appearing poles are localized except for:
\begin{align}
\label{poletypes}
  1.)&~ \text{the pole }\frac{1}{\sigma_{n}-\rho-\sum_{i=1}^{m-1}\frac{\xi_i}{2}},\notag\\
  2.)&~ \text{xor a single pole }\frac{1}{\xi_i}\text{  out of }i\in\{1,2,...,m-1\},\\
  3.)&~ \text{xor the pole }\frac{1}{\rho-\sum_{i=1}^{m-1}\frac{\xi_i}{2}-\sigma_{m+1}}.\notag
\end{align}
We choose to parametrize the $m$ localized poles in the above three cases by a parameter $\bar\rho\to0$ as follows: 
\begin{align}
  1.)&~ \rho=\bar\rho+\sigma_{m+1},~~~\xi_j=\bar\rho\,\bar\xi_j\text{ for all }j\in\{1,2,...,m-1\},\notag\\
  2.)&~ \rho=\bar\rho+\frac{1}{2}(\sigma_{m+1}+\sigma_{n}),~~~\xi_i=\sigma_{n}-\sigma_{m+1}+\bar\rho\,\bar\xi_i,~~~\text{ and }~~~\xi_j=\bar\rho\,\bar\xi_j\text{ for all }j\neq i,\\
  3.)&~ \rho=\bar\rho+\sigma_{n},~~~\xi_j=\bar\rho\,\bar\xi_j\text{ for all }j\in\{1,2,...,m-1\}.\notag
\end{align}
The new variables $\bar\xi_i$ account for the original degrees of freedom of $\xi_i$ variables at leading order after localizing $\bar\rho\to0$. Note that in all three cases we have $d\rho=d\bar\rho$, and the one pole that is not localized always directly reduces to $1/\sigma_{n,m+1}$ under $\bar\rho\to0$, which cancels the numerator in (\ref{ptlike}). In general, if we define \footnote{Note that only $m-1$ of the $\bar\sigma_q$ are now linearly independent since we have $\bar\sigma_{m}=2-\bar\sigma_1$.}
\begin{align}
\label{barsigtraf}
\bar\sigma_n=\bar\sigma_0=\bar\sigma_{m+1}\equiv 0~~~\text{and}~~~\bar\sigma_q&=1-\sum_{a=1}^{q-1}\frac{\bar\xi_a}{2}+\sum_{b=q}^{m-1}\frac{\bar\xi_b}{2}~~~\text{for}~~~~q\in\{1,2,...,m\},\\
d\bar\xi_1\wedge d\bar\xi_2\wedge...\wedge d\bar\xi_{m-1}&=2 \,d\bar\sigma_1\wedge d\bar\sigma_2\wedge...\wedge d\bar\sigma_{m-1}
\end{align}
then, for all possible pole combinations, the behavior of (\ref{ptlike}) for $\bar\rho\to0$ is parametrized as
\begin{align}
\frac{\sigma_{n,m+1}}{\sigma_{n1}\sigma_{12}...\sigma_{m,m+1}}= \frac{1}{\bar\rho^m\prod_{{a=1}\atop{a\neq r}}^{m+1}\bar\sigma_{a-1,a}}+O\left(\frac{1}{\bar\rho^{m-1}}\right),
\end{align}
where index $r\in\{1,2,...,m+1\}$ labels which one of the $m+1$ poles in the denominator of (\ref{ptlike}) is not being localized. Similarly, for all $m+1$ possible pole combinations we obtain
\begin{align}
\frac{1}{\sum_{a=1}^m\sum_{b=m+1}^n\frac{k_a\cdot k_b}{\sigma_{ab}}}=\frac{1}{\bar\rho\left(\sum_{a=1}^{r-1}\frac{k_a\cdot k_n}{\bar\sigma_a}+\sum_{b=r}^{m}\frac{k_b\cdot k_{m+1}}{\bar\sigma_b}\right)}+O(\bar\rho^0),
\end{align}
with the same index $r$. Depending on the particular value of $r$ we also get\footnote{Recall that $H$ is the derivative matrix of scattering equations. This means it is composed of elements $k_i\cdot k_j/\sigma_{ij}^2$ and their sums. While $i,j\in\{1,2,...,n\}$ initially, localizing the poles from the Parke-Taylor like factor (\ref{ptlike}) as described above removes all dependence on $\sigma_{m+1},...,\sigma_n$. This factorizes the scattering equations and their Jacobian from the remaining $(n-m)$-point amplitude.}
\begin{align}
\frac{1}{\det(H)}=\frac{\bar\rho^{2m-2}}{\det(H_r)}+O\left(\bar\rho^{2m-1}\right)~~~\text{and}~~~\text{Pf}(\psi)=\frac{1}{\bar\rho^m}\text{Pf}(\psi_r)+O\left(\frac{1}{\bar\rho^{m-1}}\right),
\end{align}
where now $H_r$ and $\psi_r$ only contain terms supported on the localized poles appearing in the Parke-Taylor like factor (\ref{ptlike}) for each $r$. It is only at this point that the scattering equations $\{h_r\}=0$, their Jacobian $1/\det(H_r)$ and all other terms become completely factorized from the remaining $(n-m)$-point amplitude $A_{n-m}$. This means $H_r$ and $\psi_r$ only depend on momenta $k_n^\mu,k_1^\mu,k_2^\mu,...,k_{m+1}^\mu$ and polarizations $\epsilon_1^\mu,\epsilon_2^\mu,...,\epsilon_{m}^\mu$, as expected. 

Plugging the above findings into (\ref{scontour}) and collecting the overall power of $\bar\rho$ we observe
\begin{align}
S_m=\sum_{r=1}^{m+1}\sum_{{\{h_r\}=0}\atop{\text{solutions}}}\oint \frac{d\bar\rho}{2\pi i} \left(\frac{1}{\bar\rho}\frac{m}{\det(H_r)}\frac{1}{\sum_{a=1}^{r-1}\frac{k_a\cdot k_n}{\bar\sigma_a}+\sum_{b=r}^{m}\frac{k_b\cdot k_{m+1}}{\bar\sigma_b}}\frac{1}{\prod_{{c=1}\atop{c\neq r}}^{m+1}\bar\sigma_{c-1,c}}\text{Pf}(\psi_r)+O(\bar\rho^0)\right),\notag
\end{align}
so that it is now trivial to compute the residues in $\bar\rho$, since for all $r$ we just have a single simple pole at $\bar\rho =0$. The result is
\begin{align}
\label{preres}
S_m=\sum_{r=1}^{m+1}\sum_{{\{h_r\}=0}\atop{\text{solutions}}}\frac{m}{\det(H_r)}\frac{1}{\sum_{a=1}^{r-1}\frac{k_a\cdot k_n}{\bar\sigma_a}+\sum_{b=r}^{m}\frac{k_b\cdot k_{m+1}}{\bar\sigma_b}}\frac{1}{\prod_{{c=1}\atop{c\neq r}}^{m+1}\bar\sigma_{c-1,c}}\text{Pf}(\psi_r).
\end{align}
Under closer inspection we note that the Pfaffian factorizes as $\text{Pf}(\psi_r)=\text{Pf}\left(\psi_{[1,r-1]}^{(n)}\right)\text{Pf}\left(\psi_{[r,m]}^{(m+1)}\right)$ with definitions (\ref{softps}), again due to trivial factorization properties of Pfaffians of block matrices with some zero blocks.

 In principle, (\ref{preres}) is already the final completely factorized result. For convenience, we can rewrite it by reassembling the Jacobian and the sum over solutions back into a shape of delta function integrations. This leads to our final general formula:\footnote{Note the convention $\text{Pf}\left(\psi_{[i,j]}^{(w)}\right)\equiv 1$ when $i>j$.}
\begin{align}
\label{softfacres}
S_m=&\sum_{r=1}^{m+1}\int d\nu_r\frac{1}{\prod_{\substack{c=1\\c\neq r}}^{m+1}\bar\sigma_{c-1,c}}\text{Pf}\left(\psi_{[1,r-1]}^{(n)}\right)\text{Pf}\left(\psi_{[r,m]}^{(m+1)}\right),\\
\label{softmeasure}
d\nu_r\equiv&~\prod_{i=1}^{m-1}d\bar\sigma_i\prod_{q=1}^{m-1}\delta(h_{q,r})\frac{2m}{\sum_{a=1}^{r-1}\frac{k_a\cdot k_n}{\bar\sigma_a}+\sum_{b=r}^{m}\frac{k_b\cdot k_{m+1}}{\bar\sigma_b}},
\end{align}
where, identifying $k_0^\mu\equiv k_n^\mu$ and keeping $\bar\sigma_0\equiv\bar\sigma_n=\bar\sigma_{m+1}= 0$ and $\bar\sigma_m=2-\bar\sigma_1$ in mind, we have
\begin{align}
\label{softscat}
h_{q,r}=\sum_{a=q}^{q+1}\sum_{\substack{b=0\\b\neq a}}^{m+1}(-1)^{a-q}\frac{k_a\cdot k_b}{\bar\sigma_{ab}}\theta_{(r-a-\frac{1}{2})(r-b-\frac{1}{2})}, 
\end{align}
with $\theta_x\equiv\theta(x)$ being the Heaviside step function. We call the constraints $h_{q,r}=0$ the soft scattering equations. The $2(j-i+1)\times 2(j-i+1)$ matrix $\psi_{[i,j]}^{(w)}$ can be written explicitly as
\begin{align}
\label{softps}
\psi_{[i,j]}^{(w)}=&\left({{A_{[i,j]}}\atop{C_{[i,j]}^{(w)}}}~{{-(C_{[i,j]}^{(w)})^T}\atop{B_{[i,j]}}}\right),~~~\text{with $(j-i+1)\times (j-i+1)$ sub-matrices}\\
A_{[i,j]}=&\left\{{{\frac{k_a\cdot k_b}{\bar\sigma_{ab}}}\atop{0}}{{;}\atop{;}}{{ a\neq b }\atop{a=b}}\right.~~,~~B_{[i,j]}=\left\{{{\frac{\epsilon_a\cdot \epsilon_b}{\bar\sigma_{ab}}}\atop{0}}{{;}\atop{;}}{{a\neq b}\atop{a=b}}\right.~~,~~C_{[i,j]}^{(w)}=\left\{\begin{matrix}
\frac{\epsilon_a\cdot k_b}{\bar\sigma_{ab}}&;\, a\neq b\\
-\frac{\epsilon_a\cdot k_w}{\bar\sigma_{a}}-\sum_{\substack{q=1\\q\neq a}}^n\frac{\epsilon_a\cdot k_q}{\bar\sigma_{aq}}&;\, a=b
\end{matrix}\right.,\notag
\end{align}
and with indices in the range $a,b\in\{i,i+1,...,j\}$. This is the final result for the $m$-soft gluon theorem in CHY formulation. We emphasize that the result is correct to leading order in $\tau\to0$. However, since (\ref{softscat}) admits different solutions of types $\bar\sigma_{a,b}=O(1)$ and $\bar\sigma_{a,b}=O(\tau)$, the integrations in (\ref{softfacres}) have to be evaluated before the result can be systematically expanded to leading order in $\tau$.

\section{Explicit examples and general pattern}
\label{examples}
In this section we work out examples for the first few soft factors $S_m$. The factors $S_1,~S_2$ and $S_3$ are obtained analytically. The factor $S_4$ (and higher) involves solutions to soft scattering equations that cannot be solved in terms of radicals, therefore we verify the validity of $S_4$ numerically. Based on the considered examples, we infer a non-trivial structural pattern for the $m$-soft factors which we conjecture to hold for any $m$.

\subsection{One-soft gluon factor $S_1$}
For $m=1$ there are no soft scattering equations (\ref{softscat}) and no delta functions to integrate. The result is just directly given by the sum over $r$ in (\ref{softfacres}):\footnote{Recall that we imply $\bar\sigma_m=2-\bar\sigma_1$, which for $m=1$ reduces to $\bar\sigma_1=1$.}
\begin{align}
\label{s1res}
S_1=2\frac{\epsilon_1\cdot k_2}{s_{12}}-2\frac{\epsilon_1\cdot k_n}{s_{1n}},
\end{align}
which clearly is the correct Weinberg soft factor.\footnote{The $s_{ij}=(k_i+k_j)^2$ is the usual Mandelstam variable.} We see that the soft factor is composed out of two pieces such as:
\begin{align}
\label{s1piece}
P^{(1)}_{1,2}\equiv 2\frac{\epsilon_1\cdot k_2}{s_{12}}.
\end{align}
 Anticipating the structure of higher $m$-soft factors, we also define
\begin{align}
\label{s0piece}
P^{(0)}_{m+1}=P^{(0)}_{n}\equiv 1.
\end{align}
Using (\ref{s0piece}) and (\ref{s1piece}) we can structurally write the Weinberg soft factor (\ref{s1res}) as
\begin{align}
\label{s1res2}
S_1&=P^{(1)}_{1,2}P^{(0)}_{n}-P^{(0)}_{2}P^{(1)}_{1,n}.
\end{align}
Based on this and further explicit results of this section, we propose in (\ref{smres}) that this structure generalizes and persists for all higher $m$-soft factors.

Restricting to four dimensions, we can convert the soft factor $S_1$ to spinor helicity formalism. We use the following standard dictionary to convert expressions of given helicity:
\begin{align}
\label{momspin}
k_i\cdot k_j = \frac{1}{2}\langle ij\rangle [ji]&,&\epsilon^{+}_i\cdot k_j =\frac{[ij]\langle jr_i\rangle}{\sqrt{2}\langle r_ii\rangle}&,&\epsilon^{-}_i\cdot k_j =\frac{\langle ij\rangle[jr_i]}{\sqrt{2}[ir_i]},\\
\label{polspin}
\epsilon^{+}_i\cdot\epsilon^{-}_j =\frac{\langle jr_i\rangle[ir_j]}{[jr_j]\langle r_ii\rangle}&,&\epsilon^{+}_i\cdot\epsilon^{+}_j =\frac{\langle r_ir_j\rangle[ji]}{\langle r_ii \rangle\langle r_jj\rangle}&,&\epsilon^{-}_i\cdot\epsilon^{-}_j =\frac{\langle ij\rangle[r_jr_i]}{[ir_i][ jr_j]},
\end{align}
where $r_i$ and $r_j$ label reference spinors assigned to spinor $i$ and $j$ respectively. With an appropriate choice of reference spinor, we see in four dimensions:
\begin{align}
S^+_1=\frac{\langle n2\rangle}{\langle n1\rangle\langle 12\rangle},
\end{align}
which is the expected familiar single soft factor in spinor helicity formalism. For real momenta, $S^-_1$ is given by complex conjugation of $S^+_1$. Here we have suppressed an overall factor of $\sqrt{2}$ in $S^+_1$ per usual spinor helicity convention.

\subsection{Two-soft gluons factor $S_2$}
For $m=2$, there is one soft scattering equation (\ref{softscat}) for each $r$, and the number of solutions organizes as follows for the different solution types and different values of $r$:
\begin{align}
\begin{array}{c|c|c|c}
\text{ solution type }&~r=1~&~r=2~&~r=3~\\ \hline
\bar\xi_1\sim O(1)    &1&1&1\\ \hline
\bar\xi_1\sim O(\tau) &1&0&1
\end{array}.
\end{align}
Adding up the contributions of all $5$ solutions and expanding to leading order in $\tau$, we obtain the following expression for $S_2$:
\begin{align}
\label{s2res}
S_2&=P^{(2)}_{1,2,3}P^{(0)}_{n}-P^{(1)}_{2,3}P^{(1)}_{1,n}+P^{(0)}_{3}P^{(2)}_{2,1,n}.
\end{align}
This agrees with the generalization (\ref{smres}) for $m=2$. The quantities $P^{(0)}_{i}$ and $P^{(1)}_{i,j}$ are given by (\ref{s0piece}), (\ref{s1piece}), and the new contribution of type $P^{(2)}_{i,j,l}$ reads:\footnote{Here, for brevity we use that $2(k_1+k_2)\cdot k_3\approx s_{123}$ at leading order in $\tau$.}
\begin{align}
\label{s2piece}
P^{(2)}_{1,2,3}=\frac{s_{13} \epsilon _1\cdot \epsilon _2}{s_{123} s_{12}}-\frac{s_{23} \epsilon _1\cdot \epsilon _2}{s_{123} s_{12}}-\frac{4 \epsilon _1\cdot
   k_3 \epsilon _2\cdot k_1}{s_{123} s_{12}}+\frac{4 \epsilon _1\cdot k_2 \epsilon _2\cdot k_3}{s_{123} s_{12}}+\frac{4 \epsilon _1\cdot k_3 \epsilon _2\cdot
   k_3}{s_{123} s_{23}}.
\end{align}
Counting the powers of $k_1$ and $k_2$ we see that this expression diverges as $\tau^{-2}$, as we expect for the two-soft gluon factor. The result (\ref{s2res}) is gauge independent and reduces to the gauge fixed result found in \cite{Volovich:2015yoa} when we select the gauge $\epsilon_2\cdot k_3=0,~\epsilon_1\cdot k_n=0$.

Restricting to four dimensions, converting to spinor helicity formalism by use of (\ref{momspin}) and (\ref{polspin}), and choosing appropriate reference spinors we get the following expression for the non-trivial helicity combination $(+-)$ after some simplification via Schouten identities:
\begin{align}
S^{+-}_2=\frac{\langle n2\rangle}{\langle n1\rangle\langle 12\rangle}\frac{[13]}{[12][23]}\left(1+\frac{\langle n1\rangle[13]\langle 32\rangle}{s_{123}\langle n2\rangle}+\frac{[1n]\langle n2\rangle[23]}{s_{n12}[13]}\right),
\end{align}
which naturally agrees with the result found in \cite{Volovich:2015yoa}. The trivial helicity combination $(++)$ reduces to the product of single soft factors $S^{++}_2={\scriptstyle\frac{\langle n3\rangle}{\langle n1\rangle\langle 12\rangle\langle 23\rangle}}$ as expected. Again, an overall factor of $(\sqrt{2})^2$ is suppressed in the above expressions per spinor helicity convention and the other helicity combinations can be obtained by complex conjugation.

We can additionally numerically test the above result in four dimensions. Making use of the GGT package provided in \cite{Dixon:2010ik} to generate explicit lower point amplitudes, we can form amplitude ratios that correspond to the soft factor in appropriate soft kinematics.\footnote{Note that there is a \texttt{Chop} command in one of the routines of the GGT package, which does not work well with soft limit numerics and therefore needs to be removed.} Keeping in mind the overall powers of $\sqrt{2}$ that are suppressed in spinor helicity, we expect to find the following relation at leading order in $\tau$:
\begin{align}
\label{numrel}
|S_m|=|\frac{(\sqrt{2})^{m}A_n(1,2,...,n)}{A_{n-m}(m+1,m+2,...,n)}|.
\end{align}
Indeed, if we generate a numeric kinematic point where $k_1^\mu,k_2^\mu$ have soft entries of order $10^{-10}$ while the rest of the momenta have hard entries of order $10^0$, we can check that i.e.
\begin{align}
 |S^{++}_2|=|\frac{2 A_6(1^+,2^+,3^+,4^+,5^-,6^-)}{A_4(3^+,4^+,5^-,6^-)}|,~~~\text{or}~~~|S^{+-}_2|=|\frac{2 A_6(1^+,2^-,3^+,4^+,5^-,6^-)}{A_4(3^+,4^+,5^-,6^-)}|,
\end{align}
hold at least to first $10$ digits, reflecting that the leading soft factor receives a first correction at the next polynomially sub-leading power in $\tau$.\footnote{To make sure that the comparison works properly, we use the same spinor conventions as the GGT package: $\lambda_i^1=\sqrt{k_i^0+k_i^3},~\lambda_i^2=(k_i^1+i k_i^2)/\sqrt{k_i^0+k_i^3}$ and $\tilde\lambda_i=(\lambda_i)^*$.} Naturally, ratios of more complicated amplitudes yield the same agreement.

\subsection{Three-soft gluons factor $S_3$}
For $m=3$, there are two soft scattering equations (\ref{softscat}) for each $r$, and the number of solutions organizes as follows for the different solution types and different values of $r$:
\begin{align}
\begin{array}{c|c|c|c|c}
\text{ solution type }&~r=1~&~r=2~&~r=3~&~r=4~\\ \hline
\bar\xi_1\sim\bar\xi_2\sim O(1)  &2&1&1&2\\ \hline
\bar\xi_i\sim O(1),~\bar\xi_j\sim O(\tau) &2&1&1&2\\ \hline
\bar\xi_1\sim\bar\xi_2\sim O(\tau) &2&0&0&2
\end{array},
\end{align}
where we imply $i\neq j$ and $i,j\in\{1,2\}$.  Adding up the contributions of all $16$ solutions and expanding to leading order in $\tau$, we obtain the following expression for $S_3$:
\begin{align}
\label{s3res}
S_3&=P^{(3)}_{1,2,3,4}P^{(0)}_{n}-P^{(2)}_{2,3,4}P^{(1)}_{1,n}+P^{(1)}_{3,4}P^{(2)}_{2,1,n}-P^{(0)}_{4}P^{(3)}_{3,2,1,n}.
\end{align}
This agrees with the generalization (\ref{smres}) for $m=3$. As before, expressions of type $P^{(0)}_{i}$, $P^{(1)}_{i,j}$ and $P^{(2)}_{i,j,l}$ are given by (\ref{s0piece}), (\ref{s1piece}) and (\ref{s2piece}), while the new contribution of type $P^{(3)}_{i,j,l,t}$ can still be analytically computed to be:\footnote{Again, we use that $2(k_1+k_2+k_3)\cdot k_4\approx s_{1234}$ and similar at leading order in $\tau$ to keep notation short.}
\begin{align}
\label{s3piece}
P^{(3)}_{1,2,3,4}&=\frac{1}{s_{12}}(w_{312}-u_{312}-u_{213}-v_{312}-v_{213})+\frac{1}{s_{23}}(w_{231}-u_{231}-u_{132}-v_{231}-v_{132})+\notag\\
&+\left(\frac{1}{s_{12}}+\frac{1}{s_{23}}\right)(u_{123}+u_{321}+v_{123}+v_{321}-w_{123})+\frac{8 \epsilon _1\cdot k_4 \epsilon _2\cdot k_4 \epsilon _3\cdot
   k_4}{s_{34} s_{234} s_{1234}}+\\
	&+\frac{8 \epsilon _1\cdot k_4 \left(\epsilon _2\cdot k_3 \epsilon _3\cdot
   k_4- \epsilon _3\cdot k_2\epsilon _2\cdot k_4\right)}{s_{23}
   s_{234} s_{1234}}+\frac{8 \left(\epsilon _1\cdot k_2 \epsilon _2\cdot k_4-
   \epsilon _2\cdot k_1\epsilon _1\cdot k_4\right) \epsilon _3\cdot k_4}{s_{12} s_{34}
   s_{1234}}+\notag\\
&+\frac{2 \epsilon _1\cdot \epsilon _2 \epsilon_3\cdot k_4}{s_{12} s_{1234}}\left(\frac{2
   s_{13}}{s_{123}}+\frac{2
   s_{14}}{s_{34}}-\frac{s_{1234}}{s_{34}}\right)+\frac{4
    \epsilon _2\cdot \epsilon _3 \epsilon _1\cdot k_4}{s_{23}
   s_{1234}}\left(\frac{s_{13}}{s_{123}}-\frac{s_{34}}{s_{234
   }}\right)+\frac{4 \epsilon _3\cdot \epsilon _1 \epsilon _2\cdot k_4}{s_{123}
   s_{1234}},\notag
\end{align}
where we used the abbreviations
\begin{align}
u_{ijl}\equiv \frac{4 \epsilon _i\cdot k_j  \epsilon_j\cdot \epsilon _l}{s_{ijl}}\left(\frac{1}{3}-\frac{ s_{l4}}{s_{ijl4}}\right),~~v_{ijl}\equiv \frac{8 \epsilon _i\cdot k_j \epsilon _j\cdot k_l \epsilon _l\cdot
   k_4}{s_{ijl} s_{ijl4}},~~w_{ijl}\equiv \frac{8 \epsilon _i\cdot k_j \epsilon _j\cdot k_4 \epsilon _l\cdot
   k_j}{s_{ijl} s_{ijl4}}.\notag
\end{align}
Counting the powers of $k_1,k_2$ and $k_3$ we see that this expression diverges as $\tau^{-3}$, as we expect for the three-soft gluon factor.

Again, we can use (\ref{momspin}) and (\ref{polspin}) to pass to spinor helicity formalism if we restrict to four dimensions. In particular, the two non-trivial independent polarization combinations are $(-+-)$ and $(+--)$. For the case $(-+-)$ we obtain, with appropriate choice of reference spinors and after some simplification via Schouten identities:
\begin{align}
S^{-+-}_{3}=&\frac{[n2]}{[n1][12]}\frac{\langle13\rangle}{\langle12\rangle\langle23\rangle}\frac{[24]}{[23][34]}\left(1-\left[\frac{\langle1n\rangle[n2]\langle23\rangle}{s_{n123}\langle13\rangle}+\frac{[2n]\langle n|k_1+k_3|2]\langle23\rangle[34]}{s_{123}s_{n123}[24]}\right.\right.\\
&~~~~~~~~~~~~~~~~~~~~~~~~~~~~~~~~~~~~~+\left.\left.\frac{[n1]\langle13\rangle[32]}{s_{123}[n2]}+\frac{\langle1n\rangle[n2]\langle23\rangle~[23]\langle3n\rangle[n4]}{\langle13\rangle s_{n12}s_{n123}[24]}+\SmallMatrix{n&\leftrightarrow&4\\1&\leftrightarrow&3}\right]\right).\notag
\end{align}
Similarly, the case $(+--)$ with an appropriate choice of reference spinors and after some simplification via Schouten identities yields
\begin{align}
S^{+--}_{3}&=\frac{\langle n2\rangle}{\langle n1\rangle\langle12\rangle}\frac{[14]}{[12][23][34]}\left(1-\frac{\langle n1\rangle[14]\langle42\rangle}{s_{1234}\langle n2\rangle}-\frac{[1n]\langle n|k_2+k_3|4]}{s_{n123}[14]}-\frac{[1n]\langle n2\rangle[23]\langle3n\rangle[n4]}{s_{n12}s_{n123}[14]}\right.\notag\\
&\left.-\frac{\langle n1\rangle[1|k_2+k_3|4\rangle[43]\langle32\rangle}{s_{123}s_{1234}\langle n2\rangle}-\frac{s_{n1}[12]\langle23\rangle[34]}{s_{123}s_{n123}[14]}+\frac{\langle n1\rangle[13]\langle32\rangle~[1n]\langle n3\rangle[34]}{\langle n2\rangle s_{123}s_{n123}[14]}\right).
\end{align}
The trivial helicity configuration $(+++)$ as expected reduces to $S^{+++}_3={\scriptstyle\frac{\langle n4\rangle}{\langle n1\rangle\langle 12\rangle\langle 23\rangle\langle 34\rangle}}$, and all other helicity configurations are obtained from the above by symmetry and complex conjugation. An overall factor of $2^{3/2}$ is suppressed in the above expressions per spinor helicity convention.

As before, (\ref{numrel}) is expected to hold. Making use of the GGT package \cite{Dixon:2010ik} to generate explicit lower point amplitudes we can form ratios that correspond to the soft factor in appropriate soft kinematics.  Generating a numeric kinematic point such that $k_1^\mu,k_2^\mu$ and $k_3^\mu$ have soft entries of order $10^{-10}$ while the rest of the momenta have hard entries of order $10^0$, we observe that i.e.
\begin{align}
|S^{-++}_3|=|\frac{2^{3/2} A_7(1^-,2^+,3^+,4^+,5^+,6^-,7^-)}{A_4(4^+,5^+,6^-,7^-)}|,~~~|S^{+-+}_3|=|\frac{2^{3/2} A_7(1^+,2^-,3^+,4^+,5^+,6^-,7^-)}{A_4(4^+,5^+,6^-,7^-)}|,~~~\text{etc.}\notag
\end{align}
hold to at least the first 10 digits, after which the first sub-leading correction in $\tau$ becomes important. Again, ratios of more complicated amplitudes yield the same agreement.

\subsection{Four-soft gluons factor $S_4$ and beyond}
For $m=4$, there are three soft scattering equations (\ref{softscat}) for each $r$, and the number of solutions organizes as follows for the different solution types and different values of $r$:
\begin{align}
\begin{array}{c|c|c|c|c|c}
\text{ solution type }&~r=1~&~r=2~&~r=3~&~r=4~&~r=5~\\ \hline
\bar\xi_1\sim\bar\xi_2\sim\bar\xi_3\sim O(1)  &5&2&1&2&5\\ \hline
\bar\xi_i\sim\bar\xi_j\sim O(1),~\bar\xi_l\sim O(\tau) &8&2&2&2&8\\ \hline
\bar\xi_i\sim O(1),~\bar\xi_j\sim\bar\xi_l\sim O(\tau) &5&2&1&2&5\\ \hline
\bar\xi_1\sim\bar\xi_2\sim \bar\xi_3\sim O(\tau) &6&0&0&0&6
\end{array},
\end{align}
where we imply $i\neq j,~i\neq l,~j\neq l$ and $i,j,l\in\{1,2,3\}$. With the generalization (\ref{smres}) in mind, we expect that the contributions for cases $r=2,3,4$ can be constructed from previously determined quantities (\ref{s1piece}), (\ref{s2piece}) and (\ref{s3piece}). That is easily verified numerically by obtaining and summing over explicit approximate solutions to the soft scattering equations (\ref{softscat}) in some example kinematics. This confirms that the structure
\begin{align}
\label{s4res}
S_4&=P^{(4)}_{1,2,3,4,5}P^{(0)}_{n}-P^{(3)}_{2,3,4,5}P^{(1)}_{1,n}+P^{(2)}_{3,4,5}P^{(2)}_{2,1,n}-P^{(1)}_{4,5}P^{(3)}_{3,2,1,n}+P^{(0)}_{5}P^{(4)}_{4,3,2,1,n}
\end{align}
continues to hold. Trying to obtain $P^{(4)}_{1,2,3,4,5}$ for $r=1$ (and $r=5$) we discover that finding the $12$ solutions of the type $\bar\xi_1\sim\bar\xi_2\sim \bar\xi_3\sim O(\tau)$ is equivalent to solving for the roots of two $6$th degree polynomials. Therefore, an analytic solution cannot be obtained in this direct fashion.

Based on the knowledge of previous analytic results found so far, we could try to infer the pole structure of all the different terms appearing in $P^{(4)}_{1,2,3,4,5}$, effectively constructing the result without solving the soft scattering equations. This works reasonably well for some of the appearing terms such as $\epsilon _1\cdot k_2 \epsilon _2\cdot k_3 \epsilon _3\cdot k_4 \epsilon _4\cdot k_5$, for which the correct contribution can be guessed (and numerically checked) to be given by:
\begin{align}
16 \frac{ \epsilon _1\cdot k_2
   \epsilon _2\cdot k_3 \epsilon _3\cdot k_4 \epsilon _4\cdot k_5}{s_{1234} s_{12345}}\left(\left(\frac{1}{s_{12}}+\frac{1}{s_{23}}\right)\frac{1}{s_{123}}+\frac{1}{s_{12}
   s_{34}}+\left(\frac{1}{s_{23}}+\frac{1}{s_{34}}\right)\frac{1}{s_{234}}\right),\notag
\end{align}
or terms like $\epsilon _1\cdot k_2 \epsilon _2\cdot k_3\epsilon _3\cdot \epsilon _4 $ with the correct guess for the contribution being:
\begin{align}
8\frac{ 
     \epsilon
   _1\cdot k_2 \epsilon _2\cdot k_3\epsilon _3\cdot \epsilon _4}{s_{1234}}\left(\frac{1}{4}-\frac{s_{45}}{s_{12345}}\right)\left(\left(\frac{1}{s_{12}}+\frac{1}{s_{23}}\right)\frac{1}{s_{123}}+\frac{1}{s_{12}
   s_{34}}+\left(\frac{1}{s_{23}}+\frac{1}{s_{34}}\right)\frac{1}{s_{234}}\right).\notag
\end{align}
However, $P^{(4)}_{1,2,3,4,5}$ also contains terms such as $\epsilon _3\cdot \epsilon _4 \epsilon _1\cdot k_2 \epsilon _2\cdot k_5$ or $\epsilon _1\cdot \epsilon _2 \epsilon _3\cdot \epsilon _4$ for which the pole structure is unclear since these patterns did not appear before. Even though an analytic solution is thus not available, we can still check numerically that (\ref{softfacres}) is correct.

Using (\ref{momspin}) and (\ref{polspin}) to pass to spinor helicity formalism in four dimensions, (\ref{numrel}) is again expected to hold. Therefore, we generate a numeric kinematic point such that $k_1^\mu,k_2^\mu,k_3^\mu$ and $k_4^\mu$ have soft entries of order $10^{-10}$ while the rest of the momenta have hard entries of order $10^0$. Now we can solve (\ref{softscat}) numerically and obtain the numeric soft factor $S_4$ as a sum over all $64$ solutions. Subsequently, making use of the GGT package \cite{Dixon:2010ik}, we can generate explicit amplitude ratios and observe that e.g.
\begin{align}
|S^{-+++}_4|=|\frac{4 A_8(1^-,2^+,3^+,4^+,5^+,6^+,7^-,8^-)}{A_4(5^+,6^+,7^-,8^-)}|,~|S^{-+-+}_4|=|\frac{4 A_8(1^-,2^+,3^-,4^+,5^+,6^+,7^-,8^-)}{A_4(5^+,6^+,7^-,8^-)}|,~\text{etc.}\notag
\end{align}
hold to at least the first 10 digits, after which the first sub-leading correction in $\tau$ becomes important. As before, ratios of more complicated amplitudes yield the same agreement.

For even higher $m$, the soft scattering equations (\ref{softscat}) become more and more complicated, so that even numeric evaluation becomes increasingly harder to do. However, in principle the $m$-soft gluon factor is always given by the CHY type expression summarized by (\ref{softfacres}), (\ref{softscat}) and (\ref{softps}), valid to leading order in $\tau$. 

\subsection{Conclusion and general structural pattern}
The above findings are of interest since they prove the existence of a universal soft factor for any number of soft adjacent gluons and in principle provide a way to calculate these soft factors in arbitrary dimension. As a byproduct we obtained an explicit analytic result for the three-soft gluon factor for arbitrary polarizations and in arbitrary dimension, which to our knowledge is a new result.\\
\indent Considering the particular results for $m=1,2,3,4$ discussed above, we can infer a generalization for the structural pattern at arbitrary $m$ to be given by:
\begin{align}
\label{smres}
S_m&=\sum_{r=1}^{m+1}(-1)^{r+1}P^{(m+1-r)}_{r,r+1,...,m,m+1}P^{(r-1)}_{r-1,r-2,...,1,n}.
\end{align}
In essence, if all soft factors $S_a$ with $a<m$ for a fixed $m$ are known, then all contributions to $S_m$ with $1<r<m+1$ are constructed from the lower point results, while the summand\footnote{Or alternatively the summand $r=m+1$, which is related by simple index exchange.} $r=1$ equals the only previously unknown contribution $P^{(m)}_{1,2,...,m,m+1}$. In general we define $P^{(0)}_{m+1}=P^{(0)}_{n}\equiv 1$ and
\begin{align}
\label{Pdef}
P^{(i)}_{1,2,...,i,i+1}=&\int d\nu_1\frac{1}{\prod_{c=2}^{i+1}\bar\sigma_{c-1,c}}\text{Pf}\left(\psi_{[1,i]}^{(i+1)}\right).
\end{align}
In this sense, it suffices to evaluate only the $r=1$ summand of (\ref{softfacres}) to obtain all new information at a given $m$.\footnote{There seems to be no obstruction to assuming that a similar pattern should appear for soft theorems e.g. in the other theories discussed below as well, where appropriate.}

The above conjecture is inferred empirically, and it seems to be highly non-trivial to demonstrate the factorization of each summand of (\ref{softfacres}) into (\ref{smres}) analytically. While the structure of the Pfaffian admits such a factorization, the Parke-Taylor like factor as well as the multiplicative term remaining from the contour deformation in $\rho$ are not convenient. This implies the necessity of a transformation along the lines of (\ref{deltatrafo}) with a non-trivial Jacobian, which is not easily guessed. We leave a general proof of the conjecture (\ref{smres}), (\ref{Pdef}) to future work.

\section{Multi-soft factors in other theories}
\label{othertheories}
It is possible to directly apply the procedure described above to several other theories in CHY formulation. An important feature that largely governs the computations is the presence of at least one Parke-Taylor factor
\begin{align}
\mathcal{C}\equiv\frac{1}{\sigma_{12}\sigma_{23}...\sigma_{n1}}
\end{align}
in the CHY integrand of the amplitude, such that the amplitude in question is color ordered. The theories considered in this section have this same feature. As further building blocks we will require the sub-matrix $A$ defined in (\ref{psimat}), the matrix $\Psi_{n+1,n+2,...,n+q}^{n+1,n+2,...,n+q}$ which is (\ref{psimat}) with rows and columns $n+1,n+2,...,n+q$ dropped, and the matrix
\begin{align}
\label{chimatrix}
\chi=\left\{{{\frac{\delta^{I_a,I_b}}{\sigma_{ab}}}\atop{0}}{{;}\atop{;}}{{a\neq b}\atop{a=b}}\right. ,
\end{align}
where $I_a,I_b$ are some internal space indices for scalar fields involved in the scattering process \cite{Cachazo:2014xea}. Since these indices have no non-trivial effect on the momentum dependence of soft factors, we will consider the simplest case where $I_a=I_b$ for all particle labels $a,b$ , such that $\delta^{I_a,I_b}=1$.

\subsection{Multi-soft factors in bi-adjoint scalar $\phi^3$ theory}
The CHY formula for tree level scattering in bi-adjoint scalar $\phi^3$ theory can be written as (\ref{ampl}) \cite{Cachazo:2013hca} with $\mathcal{I}_n^{YM}$ replaced by
\begin{align}
\mathcal{I}_n^{\phi^3}=\mathcal{C}^2.
\end{align}
Starting with this integrand, the considerations in sections \ref{CHYsoft} and \ref{factorization} go through in the same manner, such that we are left with the following general expression for the $m$-soft scalar factor with particles $1,2,...,m$ going soft:
\begin{align}
\label{SmScalar}
S^{\phi^3}_m=\sum_{r=1}^{m+1}\int d\nu_r \frac{1}{\prod_{\substack{c=1\\c\neq r}}^{m+1}(\bar\sigma_{c-1,c})^2},
\end{align}
with $d\nu_r$ given in (\ref{softmeasure}), and the identification $\sigma_0\equiv\sigma_n$. As in the gluon case, the soft scattering equations contained in $d\nu_r$ can be explicitly solved for the cases $m=1,2,3$, with exactly the same solutions. At leading order in the soft limit this leads to
\begin{align}
S^{\phi^3}_1=&\frac{1}{k_n\cdot k_1}+\frac{1}{k_1\cdot k_2},\\
S^{\phi^3}_2=&\frac{1}{k_1\cdot k_2}\left(\frac{1}{k_n\cdot (k_1+k_2)}+\frac{1}{(k_1+k_2)\cdot k_3}\right),\\
S^{\phi^3}_3=&\frac{2}{s_{123}}\left(\frac{1}{k_1\cdot k_2}+\frac{1}{k_2\cdot k_3}\right)\left(\frac{1}{k_n\cdot (k_1+k_2+k_3)}+\frac{1}{(k_1+k_2+k_3)\cdot k_4}\right).
\end{align}
It is worth noticing that all contributions to the soft factors at leading order in the soft limit are due to the two summands $r=1$ and $r=m+1$ only, while the intermediate summands are sub-leading. As before, the general expression $S^{\phi^3}_m$ can be used to evaluate $S^{\phi^3}_4$ and higher soft factors numerically. We tested the results numerically against amplitude ratios in CHY formulation and found agreement.

\subsection{Multi-soft factors in Yang-Mills-scalar theory}
The CHY formula for tree level scattering in Yang-Mills-scalar theory is (\ref{ampl}) with $\mathcal{I}_n^{YM}$ replaced by
\begin{align}
\mathcal{I}_n^{YMS}=2\,\mathcal{C}\,\text{Pf}(\chi)\frac{(-1)^{i+j}}{\sigma_{ij}}\text{Pf}(\Psi_{i,j,n+1,n+2,...,n+q}^{i,j,n+1,n+2,...,n+q}),
\end{align}
where matrix $\chi$ is $q\times q$ dimensional (\ref{chimatrix}), and $1\leq i<j\leq n$ can be selected arbitrarily \cite{Cachazo:2014xea}. This corresponds to the first $q$ of the scattering particles being scalars and the remaining $n-q$ being gluons.\\
Starting with this integrand, the considerations in sections \ref{CHYsoft} and \ref{factorization} go through in the same manner. Soft gluon factors in this theory are exactly the same as in pure Yang-Mills. The general expression for the $m$-soft scalar factor with particles $1,2,...,m$ going soft amounts to:\footnote{Again, we introduce the convention $\text{Pf}(\chi_{[i,j]})=\text{Pf}(A_{[i,j]})\equiv 1$ when $i>j$.}
\begin{align}
\label{SmYMS}
S^{YMS}_m=\sum_{r=1}^{m+1}\int d\nu_r \frac{1}{\prod_{\substack{c=1\\c\neq r}}^{m+1}\bar\sigma_{c-1,c}}\text{Pf}(\chi_{[1,r-1]})\text{Pf}(\chi_{[r,m]})\text{Pf}(A_{[1,r-1]})\text{Pf}(A_{[r,m]}),
\end{align}
with $d\nu_r$ given in (\ref{softmeasure}), and the identification $\sigma_0\equiv\sigma_n$. The matrix $A_{[i,j]}$ was defined in (\ref{softps}), and the matrix $\chi_{[i,j]}$ relates to $\chi$ in (\ref{chimatrix}) the same as $A_{[i,j]}$ relates to $A$ in (\ref{psimat}). As in the gluon case, the soft scattering equations contained in $d\nu_r$ can be explicitly solved for the cases $m=1,2,3$, with exactly the same solutions. However, since $\text{Pf}(\chi_{[i,j]})$ vanishes when $\chi_{[i,j]}$ is of odd dimension, only soft factors with an even number $m$ of soft scalars are non-zero and only summands of odd $r$ contribute. At leading order in the soft limit this leads to
\begin{align}
S^{YMS}_2=&\frac{1}{2k_1\cdot k_2}\left(\frac{k_n\cdot (k_2-k_1)}{k_n\cdot (k_1+k_2)}+\frac{(k_1-k_2)\cdot k_3}{(k_1+k_2)\cdot k_3}\right).
\end{align}
This agrees with the result in \cite{Cachazo:2015ksa}. As before, the general expression $S^{YMS}_m$ can be used to evaluate $S^{YMS}_4$ and higher soft factors numerically. We tested the results numerically against amplitude ratios in CHY formulation and found agreement.

\subsection{Multi-soft factors in non-linear sigma model }
The CHY formula for tree level scattering in non-linear sigma model is (\ref{ampl}) with $\mathcal{I}_n^{YM}$ replaced by
\begin{align}
\mathcal{I}_n^{NLSM}=\mathcal{C}\frac{4}{(\sigma_{ij})^2}\text{Pf}(A_{i,j}^{i,j})^2,
\end{align}
where $A_{i,j}^{i,j}$ is the matrix $A$ defined in (\ref{psimat}) with rows and columns $i,j$ removed, and $1\leq i<j\leq n$ can be selected arbitrarily \cite{Cachazo:2014xea}.\\
Starting with this integrand, the considerations in sections \ref{CHYsoft} and \ref{factorization} go through in the same manner. The general expression for the $m$-soft factor with particles $1,2,...,m$ going soft amounts to:
\begin{align}
\label{SmNLSM}
S^{NLSM}_m=\sum_{r=1}^{m+1}\int d\nu_r \frac{1}{\prod_{\substack{c=1\\c\neq r}}^{m+1}\bar\sigma_{c-1,c}}\text{Pf}(A_{[1,r-1]})^2\text{Pf}(A_{[r,m]})^2,
\end{align}
with $d\nu_r$ given in (\ref{softmeasure}), and the identification $\sigma_0\equiv\sigma_n$. The matrix $A_{[i,j]}$ was defined in (\ref{softps}). As in the gluon case, the soft scattering equations contained in $d\nu_r$ can be explicitly solved for the cases $m=1,2,3$, with exactly the same solutions. However, since $\text{Pf}(A_{[i,j]})$ vanishes when $A_{[i,j]}$ is of odd dimension, only soft factors with an even number $m$ of soft particles are non-zero and only summands of odd $r$ contribute. At leading order in the soft limit this leads to
\begin{align}
S^{NLSM}_2=&\frac{1}{2}\left(\frac{k_n\cdot (k_2-k_1)}{k_n\cdot (k_1+k_2)}+\frac{(k_1-k_2)\cdot k_3}{(k_1+k_2)\cdot k_3}\right).
\end{align}
This agrees with the result in \cite{Cachazo:2015ksa}. As before, the general expression $S^{NLSM}_m$ can be used to evaluate $S^{NLSM}_4$ and higher soft factors numerically. We tested the results numerically against amplitude ratios in CHY formulation and found agreement. Additionally, our $S^{NLSM}_4$ numerically agrees with the result found in \cite{Du:2016njc}.\footnote{Note a typo in eq. (4.10) of \cite{Du:2016njc}: The numerator of last expression on the first line should involve $q_5\cdot k_1$ instead of $q_4\cdot k_1$.}

\appendix
\section{CSW recursion for multi-gluon soft-factors in four dimensions}
\label{CSWrecur}
As an alternative to the construction rules presented in \cite{Georgiou:2015jfa}, we can set up a CSW type recursion \cite{Cachazo:2004kj} for the $m$-soft factors in four dimensions as follows. We start with the amplitude $A^{(m)}(k_n^{+1},k_1^{h_1},...,k_m^{h_m},k_{m+1}^{+1})$, where $k_i^{h_i}$ denotes the external momentum of the $i$-th particle with helicity $h_i\in\{+1,-1\}$. Here we have cyclically rotated the $n$-th position to be the first, and suppressed all entries $k_j^{h_j}$ with $m+1<j<n$ since they do not enter the soft factor that we want to extract from this amplitude. Since the helicities of particle $n$ and $m+1$ do not enter the soft factor, we can choose these helicities to be $+$ without loss of generality. The superscript $(m)$ keeps track of the number of adjacent external momenta that are taken soft. 

In order to obtain the soft factor from CSW recursion, we have to generate all possible diagrams in MHV expansion. To do this recursively, we introduce the following two functions:
\begin{align}
\label{funS}
&\bold{S}\left(A^{(m)}(k_{q_1}^{h_{q_1}},k_{q_2}^{h_{q_2}},...,k_{q_l}^{h_{q_l}})\right)=\\
&=\left\{\begin{matrix}
 {\substack{\displaystyle\sum_{\nu=\pm 1}\sum_{i=1}^{l-1}\sum_{{\substack{j=i+1\\j-i<l-1}}}^{l}\bold{H}\left(A_{j-i+2}(k_{q_i}^{h_{q_i}},...,k_{q_j}^{h_{q_j}},k_{p(q_i,...,q_j)}^{-\nu})\right)\frac{1}{P^2_{q_i,...,q_j}}\times\\\displaystyle~~~~~~~~~~\times\bold{S}\left(A^{(m)}(k_{q_{1}}^{h_{q_{1}}},...,k_{q_{i-1}}^{h_{q_{i-1}}},k_{p(q_i,...,q_j)}^{+\nu},k_{q_{j+1}}^{h_{q_{j+1}}},...,k_{q_l}^{h_{q_l}})\right)}}&~~~;~~~\displaystyle\text{ if }\sum_{a=1}^lh_{q_a}<l,\\
~&~\\
\displaystyle A^{(m)}(k_{q_1}^{h_{q_1}},k_{q_2}^{h_{q_2}},...,k_{q_l}^{h_{q_l}})&~~~;~~~\displaystyle~\text{ otherwise, }~
\end{matrix}\right.\notag
\end{align}
as well as, making use of $\mu(x)\equiv\text{mod}(x-1,l)+1$, the function:
\begin{align}
\label{funH}
&\bold{H}\left(A_l(k_{q_1}^{h_{q_1}},k_{q_2}^{h_{q_2}},...,k_{q_l}^{h_{q_l}})\right)=\\
&=\left\{\begin{matrix}
 {\substack{\displaystyle\sum_{i=1}^{l}\sum_{{\substack{j=i+1}}}^{i+l-3}\bold{H}\left(A_{j-i+2}(k_{q_{\mu(i)}}^{h_{q_{\mu(i)}}},...,k_{q_{\mu(j)}}^{h_{q_{\mu(j)}}},k_{p(q_{\mu(i)},...,q_{\mu(j)})}^{-1})\right)\frac{1}{P^2_{q_{\mu(i)},...,q_{\mu(j)}}}\times\\\displaystyle~~~~~~~~~~\times\bold{H}\left(A_{l+i-j}(k_{q_{\mu(j+1)}}^{h_{q_{\mu(j+1)}}},...,k_{q_{\mu(l+i-1)}}^{h_{q_{\mu(l+i-1)}}},k_{p(q_{\mu(j+1)},...,q_{\mu(l+i-1)})}^{+1})\right)}}&~~;~~\displaystyle\text{ if }\sum_{a=1}^lh_{q_a}<l-4,\\
~&~\\
\displaystyle A_l(k_{q_1}^{h_{q_1}},k_{q_2}^{h_{q_2}},...,k_{q_l}^{h_{q_l}})&;~~~\,~\displaystyle~\text{ otherwise. }~
\end{matrix}\right.\notag
\end{align}
We supplement the above functions with the following resolution properties:
\begin{align}
p(i,...,j,p(a,...,b),u,...,v)=&p(i,...,j,a,...,b,u,...,v),\\
p(i,...,j,r,a,...,b,r,u,...,v)=&p(i,...,j,a,...,b,u,...,v),\\
P^2_{i,...,j,p(a,...,b),u,...,v}=&P^2_{i,...,j,a,...,b,u,...,v},\\
P^2_{i,...,j,r,a,...,b,r,u,...,v}=&P^2_{i,...,j,a,...,b,u,...,v},
\end{align}
which ensure that the explicit propagator momenta always are properly resolved in terms of external momenta. Naturally, the order of indices $i,...,j$ appearing in $p(i,...,j)$ and $P^2_{i,...,j}$ is irrelevant and can be assumed to be sorted to make it easier to identify and group together identical expressions.

It is important to note that the sums in the functions (\ref{funS}) and (\ref{funH}) may contain summands that immediately vanish due to trivial helicity configurations of sub-amplitudes involved that enter the $\bold{H}$ function.\footnote{By trivial helicity configuration we mean amplitudes with none, or only one negative helicity gluon, as well as amplitudes with none, or only one positive helicity gluon (special care is required for $3$-point amplitudes due to special kinematics).} Setting such summands to zero directly without allowing for any recursion depth in such terms greatly speeds up the calculation.

Recursion by means of (\ref{funS}) and (\ref{funH}) with the above supplements will generate all possible diagrams in MHV expansion that contribute to leading order in the soft limit. However, the simple summation employed here comes at the expense of multiple counting for some of the resulting diagrams. The easiest way to remove the over-counting is to simply set the integer coefficient in front of each overall summand to $1$ after the recursion has been completed and all terms have been properly grouped together:
\begin{align}
\bold{S}'\equiv\bold{S}\text{ with multiplicity of each overall summand set to }1,
\end{align}
which implies that invariance of amplitudes under cyclic permutation of external legs is used to identify and group together equivalent terms in the expansion. This, as well as the entire recursive procedure, can be easily automated i.e. in \textit{Mathematica}, such that the $m$-soft factor $S_m$ for any helicity configuration is automatically generated by the input:
\begin{align}
S_m=\bold{S}'\left(A^{(m)}(k_{n}^{+1},k_{1}^{h_{1}},...,k_{m}^{h_{m}},k_{q_{m+1}}^{+1})\right).
\end{align}
Finally, to evaluate the soft factor we use the substitutions
\begin{align}
\label{subhard}
&A^{(m)}(k_{q_1}^{+1},k_{q_2}^{+1},...,k_{q_l}^{+1})\to\frac{\langle n-1,n\rangle\langle n,m+1\rangle\langle m+1,m+2\rangle}{\langle n-1,q_1\rangle\left(\prod_{i=1}^{l-1}\langle q_{i},q_{i+1}\rangle\right)\langle q_l,m+2\rangle},\\
\label{subsoft}
&A_l(k^{+1}_{q_1},...,k_{q_{i-1}}^{+1},k_{q_i}^{-1},k_{q_{i+1}}^{+1},...,k_{q_{j-1}}^{+1},k_{q_j}^{-1},k_{q_{j+1}}^{+1},...,k^{+1}_{q_l})\to\frac{\langle q_i,q_j\rangle^4}{\langle q_l,q_1\rangle\prod_{i=1}^{l-1}\langle q_{i},q_{i+1}\rangle},
\end{align}
where entries like $|p(i,...,j)\rangle$ are evaluated by the usual CSW replacement $P_{i,...,j}|X]$ with reference spinor $|X]$. Superficially, due to (\ref{subhard}) it might seem that the soft factor depends on $(n-1)$-st and $(m+2)$-nd external momentum as well. However, just as in \cite{Georgiou:2015jfa}, this dependence always cancels out upon the CSW replacement of the shifted spinors at leading order in $\tau$.

We have tested the above recursive procedure for soft factors $S_1,~S_2,...,S_7$ with various helicity configurations against appropriate amplitude ratios obtained from the GGT package \cite{Dixon:2010ik}, and found numerical agreement at leading order in $\tau$. For example, our recursion takes about two minutes to generate the $2277$ different analytic terms in the $S_7^{------+}$ soft factor. If required, a trivial further expansion in $\tau$ can be used to isolate leading terms only.

\section{Four-soft gluons from BCFW}
\label{foursoftBCFW}
Naturally, it is also possible to apply BCFW recursion relations \cite{Britto:2004ap} to compute higher soft factors. Here we demonstrate the four-soft gluon calculation. We pick gluons $1,2,3,4$ to be soft and perform a $[23\rangle$ BCFW shift, so that $2\to\hat 2$ and $3\to\hat 3$ with
\begin{align}
|\hat 2\rangle=|2\rangle~~~,~~~|\hat 2]=|2]+z|3]~~~,~~~|\hat 3\rangle=|3\rangle - z |2\rangle~~~,~~~|\hat 3]=|3].
\end{align}
It is trivial to see that under this shift only the following four diagrams could possibly contribute to the leading soft factor with any helicity configuration:
\begin{align}
S_{4,A}=&A_4(n,1,\hat 2,-\hat P_{n12})\frac{1}{s_{n12}}S_2(\hat P_{n12},\hat 3,4,5),\\
S_{4,B}=&A_3(1,\hat 2,-\hat P_{12})\frac{1}{s_{12}}S_3(n,\hat P_{12},\hat 3,4,5),\\
S_{4,C}=&A_4(-\hat P_{345},\hat 3,4,5)\frac{1}{s_{345}}S_2(n,1,\hat 2,\hat P_{345}),\\
S_{4,D}=&A_3(-\hat P_{34},\hat 3,4)\frac{1}{s_{34}}S_3(n,1,\hat 2,\hat P_{34},5),
\end{align}
while the complete four-soft gluon factor is given by 
\begin{align}
S_4=S_{4,A}+S_{4,B}+S_{4,C}+S_{4,D}
\end{align}
in each case. Here, $A_3,A_4$ are mostly-soft-leg sub-amplitudes factored by BCFW, and $S_2,S_3$ are two- and three-soft gluon factors that are extracted from the mostly-hard-leg sub-amplitudes factored by BCFW. The usual on-shell constraints $\hat P^2_{\cdots}=0$ provide the following $z$ values to leading order in the soft limit:\footnote{We use the convention $s_{ij}=\langle ij\rangle[ji]$, which with our spinor contraction conventions ($\langle ij\rangle=\lambda_i^1\lambda_j^2-\lambda_i^2\lambda_j^1$ and $[ij]=\tilde\lambda_i^2\tilde\lambda_j^1-\tilde\lambda_i^1\tilde\lambda_j^2$) corresponds to $(+,-,-,-)$ Minkowski metric signature.}
\begin{align}
z_A=\frac{-s_{n12}}{\langle 2n\rangle[n3]}~~~,~~~z_B=-\frac{[12]}{[13]}~~~,~~~z_C=\frac{s_{345}}{\langle25\rangle [53]}~~~,~~~z_D=\frac{\langle34\rangle}{\langle24\rangle}.
\end{align}
In case when all four soft gluons have the same helicity, the four-soft factor trivially reduces to a product of consecutive soft factors. In the following, we specify explicit helicity configurations and obtain the results for all analytically distinct non-trivial helicity configurations.

\paragraph{Helicity configuration $(-+++)$:}$~$\\
For the helicity configuration of soft gluons $(1^-,2^+,3^+,4^+)$ we find:
\begin{align}
S_{4,A}^{-+++}=&~\frac{[3\, n]^3 \langle 1\, n\rangle ^3 \langle 5\, n\rangle }{s_{n12} s_{n123} \langle 1\, 2\rangle  \langle 4\, 5\rangle  \langle n|k_{12}|3] \langle 4|k_{n123}k_{n1}|2\rangle },\\
S_{4,B}^{-+++}=&~\frac{[2\, 3]^3 \langle n\, 5\rangle }{s_{123} [1\, 2] \langle 4\, 5\rangle  \langle 4|k_{23}|1] \langle n|k_{12}|3]},\\
S_{4,C}^{-+++}=&~0,\\
S_{4,D}^{-+++}=&~\frac{\langle n\, 5\rangle  \langle 4|k_{23}|n]{}^3}{\langle 2\, 3\rangle  \langle 3\, 4\rangle  \langle 4\, 5\rangle  [n\, 1] \langle 4|k_{23}|1] \langle 4|k_{123}|n] \langle 2|k_{n1}k_{n123}|4\rangle }+\\
&+\frac{\langle 1\, 5\rangle ^3 [n\, 5]}{s_{12345} \langle 1\, 2\rangle  \langle 2\, 3\rangle  \langle 3\, 4\rangle  \langle 4\, 5\rangle  \langle 5|k_{1234}|n]}+\frac{\langle n\, 5\rangle  \langle 1|k_{234}|n]{}^3}{s_{1234} s_{\text{n1234}} \langle 1\, 2\rangle  \langle 2\, 3\rangle  \langle 3\, 4\rangle  \langle 4|k_{123}|n] \langle 5|k_{1234}|n]}.\notag
\end{align}
To see that the diagram $C$ is zero, we use the fact that the soft factor is independent of the helicity of particle $5$, thus we can choose it to be $5^+$ which leads to no non-vanishing helicity configurations for $A_4$. In all other diagrams only one helicity configuration is non-vanishing. We tested the above result numerically against amplitude ratios and found agreement.

\paragraph{Helicity configuration $(+-++)$:}$~$\\
For the helicity configuration of soft gluons $(1^+,2^-,3^+,4^+)$ we find:
\begin{align}
&S_{4,A}^{+-++}=\frac{[3\, n]^3 \langle 2\, n\rangle ^4 \langle 5\, n\rangle }{s_{\text{n12}} s_{\text{n123}} \langle 1\, 2\rangle  \langle 4\, 5\rangle  \langle 1\, n\rangle  \langle n|k_{12}|3] \langle 4|k_{\text{n123}}k_{\text{n1}}|2\rangle },\\
&S_{4,B}^{+-++}=\frac{[1\, 3]^4 \langle n\, 5\rangle }{s_{123} [1\, 2] [2\, 3] \langle 4\, 5\rangle  \langle 4|k_{23}|1] \langle n|k_{12}|3]},\\
&S_{4,C}^{+-++}=~0,\\
&S_{4,D}^{+-++}=\frac{\langle 5\, n\rangle }{\langle 2\, 3\rangle  \langle 3\, 4\rangle  \langle 5|k_{234}|1]}\left(\frac{[1\, 5]^3 \langle 2\, 5\rangle ^4}{s_{12345} s_{2345} \langle 4\, 5\rangle  \langle 2|k_{345}k_{12345}|n\rangle }+\frac{\langle 2|k_{34}|1]{}^4}{s_{1234} s_{234} \langle 4|k_{23}|1] \langle 2|k_{34}k_{1234}|n\rangle }\right)\notag\\
&+\frac{\langle 2\, n\rangle ^3}{\langle 1\, 2\rangle  \langle 2\, 3\rangle  \langle 3\, 4\rangle  \langle 1\, n\rangle  \langle 2|k_{\text{n1}}k_{\text{n1234}}|5\rangle }\left(\frac{\langle 2\, 5\rangle ^3 [n\, 5]}{\langle 4\, 5\rangle  \langle 2|k_{345}k_{12345}|n\rangle }+\frac{\langle 2\, n\rangle  \langle 5\, n\rangle  \langle 2|k_{34}|n]{}^3}{s_{\text{n1234}} \langle 2|k_{\text{n1}}k_{\text{n123}}|4\rangle  \langle n|k_{1234}k_{34}|2\rangle }\right).
\end{align}
Diagram $C$ vanishes the same way as described above. In all other diagrams again only one helicity configuration is non-vanishing. We tested the above result numerically against amplitude ratios and found agreement.

\paragraph{Helicity configuration $(+--+)$:}$~$\\
For the helicity configuration of soft gluons $(1^+,2^-,3^-,4^+)$ we find:
\begin{align}
&S_{4,A}^{+--+}=\frac{\langle 2\, n\rangle ^3}{s_{\text{n12}} \langle 1\, 2\rangle  \langle 1\, n\rangle  \langle n|k_{12}|3] \langle 2|k_{\text{n1}}k_{\text{n1234}}|5\rangle }\left(\frac{[4\, n]^3 \langle 2\, n\rangle  \langle 5\, n\rangle }{s_{\text{n1234}} [3\, 4]}+\right.\\
&~~~~~~~~~~+\left.\frac{[5\, n] \langle 2|k_{\text{n1}}k_{\text{n123}}|5\rangle {}^3}{\langle 4\, 5\rangle  \langle 2|k_{\text{n1}}k_{\text{n123}}|4\rangle  \left(s_{345} [3\, n] \langle 2\, n\rangle +s_{\text{n12}} [3\, 5] \langle 2\, 5\rangle \right)}\right)\notag\\
&S_{4,B}^{+--+}=\frac{\langle n|k_{23}|1]{}^3 }{s_{123} [1\, 2] [2\, 3] \langle n|k_{12}|3] \left(\langle 5|k_{1234}|n] \langle
   n|k_{23}|1]-\langle 5\, n\rangle  [1|k_{23}k_{123}|n]\right)}\times\\
&~~~~~~~~~~\times\left(\frac{[4\, n]^3 \langle 5\, n\rangle  \langle n|k_{23}|1]}{s_{\text{n123}} s_{\text{n1234}} \langle n|k_{123}|4]}+\frac{[5\, n] \langle 5|k_{23}|1]{}^3}{\langle 4\, 5\rangle \langle 4|k_{23}|1] \left([4\, 5]  \langle 5\, n\rangle  \langle 4|k_{23}|1]+  \langle 5|k_{23}|1]  \langle n|k_{1234}|5]\right)}\right)\notag\\
&~~~~~~~~~~+\frac{1}{[1\, 2] [2\, 3] \langle 5|k_{234}|1]}\left(\frac{[1\, 4]^4 \langle 5\, n\rangle }{s_{1234} [3\, 4] \langle n|k_{123}|4]}+\right.\notag\\
&~~~~~~~~~~\left.+\frac{[1\, 5]^3 \langle n\, 5\rangle  \langle 5|k_{23}|1]{}^4}{s_{12345} \langle 4\, 5\rangle  \langle 4|k_{23}|1] [1|k_{2345}k_{45}|3] \left([4\, 5] \langle 5\, n\rangle  \langle 4|k_{23}|1]+\langle 5|k_{23}|1] \langle n|k_{1234}|5]\right)}\right),\notag\\
&S_{4,C}^{+--+}=\frac{[4\, 5]^3 \langle 2\, 5\rangle ^3 }{s_{345} [3\, 4] \langle 2|k_{34}|5] \langle 2|k_{345}k_{12345}|n\rangle }\left(\frac{[1\, 5]^3 \langle 2\, 5\rangle  \langle n\, 5\rangle }{s_{12345} s_{2345} [1|k_{2345}k_{45}|3]}+\right.\\
&~~~~~~~~~~\left.+\frac{[n\, 5] \langle 2\, n\rangle ^3}{\langle 1\, 2\rangle  \langle 1\,
   n\rangle  \left(s_{345} [3\, n] \langle 2\, n\rangle +s_{\text{n12}} [3\, 5] \langle 2\, 5\rangle \right)}\right),\notag\\
&S_{4,D}^{+--+}=\frac{\langle 2\, 3\rangle ^3}{s_{234} \langle 3\, 4\rangle  \langle 4|k_{23}|1] \langle n|k_{1234}|5]}\left(\frac{[n\, 5] \langle n|k_{234}|1]{}^3}{s_{1234} s_{\text{n1234}} \langle 2|k_{34}k_{1234}|n\rangle }+\frac{[1\, 5]^3 \langle n\, 5\rangle }{s_{12345} \langle 2|k_{34}|5]}\right)\\
&~~~~~~~~~~+\frac{\langle 2\, 3\rangle ^3 [n\, 5] \langle 2\, n\rangle ^3}{\langle 1\, 2\rangle  \langle 3\, 4\rangle  \langle 1\, n\rangle  \langle 2|k_{34}|5] \langle 4|k_{\text{n123}}k_{\text{n1}}|2\rangle  \langle 2|k_{34}k_{1234}|n\rangle }.\notag
\end{align}
In all diagrams again only one helicity configuration is non-vanishing. We tested the above result numerically against amplitude ratios and found agreement.

\paragraph{Helicity configuration $(--++)$:}$~$\\
For the helicity configuration of soft gluons $(1^-,2^-,3^+,4^+)$ we find:
\begin{align}
&S_{4,A}^{--++}=\frac{\langle 1\, 2\rangle ^3 [3\, n]^3 \langle 5\, n\rangle }{s_{\text{n12}} s_{\text{n123}} \langle 4\, 5\rangle  \langle 1\, n\rangle  \langle n|k_{12}|3] \langle 4|k_{\text{n123}}k_{\text{n1}}|2\rangle },\\
&S_{4,B}^{--++}=\frac{1}{s_{123} [1\, 2] [2\, 3] \langle 4|k_{23}|1] \langle 5|k_{1234}|n]}\left(\frac{\langle n\, 5\rangle  [n|k_{1234}k_{12}|3]{}^3}{s_{1234} s_{\text{n1234}} \langle 4|k_{123}|n]}+\frac{[5\, n] \langle 5|k_{12}|3]{}^3}{s_{12345} \langle 4\, 5\rangle }\right)\\
&~~~~~~~~~~+\frac{[3\, n]^3 \langle 5\, n\rangle }{s_{\text{n123}} [1\, 2] [2\, 3] \langle 4\, 5\rangle  [1\, n] \langle 4|k_{123}|n]},\notag\\
&S_{4,C}^{--++}=~0,\\
&S_{4,D}^{--++}=\frac{[n\, 5]}{\langle 2\, 3\rangle  \langle 3\, 4\rangle  \langle 5|k_{234}|1]}\left(\frac{\langle 2\, 5\rangle ^3}{s_{2345} \langle 4\, 5\rangle  [1\, n]}+\frac{\langle 2|k_{34}k_{1234}|5\rangle {}^3}{s_{1234} s_{12345} s_{234} \langle 4|k_{23}|1] \langle 5|k_{1234}|n]}\right)\\
&~~~~~~~~~~+\frac{\langle n\, 5\rangle  \langle 2|k_{34}|n]{}^3}{s_{234} s_{\text{n1234}} \langle 2\, 3\rangle  \langle 3\, 4\rangle  [1\, n] \langle 4|k_{23}|1] \langle 5|k_{1234}|n]}.\notag
\end{align}
Diagram $C$ vanishes the same way as described above. In all other diagrams again only one helicity configuration is non-vanishing. We tested the above result numerically against amplitude ratios and found agreement.

\paragraph{Helicity configuration $(+-+-)$:}$~$\\
For the helicity configuration of soft gluons $(1^+,2^-,3^+,4^-)$ we find:
\begin{align}
&S_{4,A}^{+-+-}=\frac{[3\, n]^3 \langle 2\, n\rangle ^4}{s_{\text{n12}} \langle 1\, 2\rangle  \langle 1\, n\rangle  [5|k_{\text{n1234}}k_{\text{n12}}|3] \langle n|k_{12}|3]}\times\\
&~~~~~~~~~~\times\left(\frac{[3\, n] [5\, n] \langle 4\, n\rangle ^3}{s_{\text{n123}} s_{\text{n1234}} \langle 4|k_{\text{n123}}k_{\text{n1}}|2\rangle }+\frac{[3\, 5]^3 \langle n\, 5\rangle }{[3\, 4] [4\, 5] \left(s_{345} [3\, n] \langle 2\, n\rangle +s_{\text{n12}} [3\, 5] \langle 2\, 5\rangle \right)}\right),\notag\\
&S_{4,B}^{+-+-}=\frac{[1\, 3]^4 [3\, 5]^3 \langle n\, 5\rangle }{[1\, 2] [2\, 3] [3\, 4] [4\, 5] [3|k_{12}k_{1234}|5] [1|k_{2345}k_{45}|3] \langle n|k_{12}|3]}+\\
&~~~~~~~~~~+\frac{[1\, 3]^4}{s_{123} [1\, 2] [2\, 3] \langle 4|k_{23}|1] \langle n|k_{1234}|5]}\left(\frac{\langle 5\, n\rangle  \langle 4|k_{123}|5]{}^3}{s_{1234} s_{12345} [3|k_{12}k_{1234}|5]}+\frac{[5\, n] \langle 4\, n\rangle ^3}{s_{\text{n1234}} \langle n|k_{12}|3]}\right),\notag\\
&S_{4,C}^{+-+-}=\frac{[3\, 5]^4 \langle 2\, 5\rangle ^3}{s_{345} [3\, 4] [4\, 5] \langle 2|k_{34}|5] \langle n|k_{12345}k_{345}|2\rangle }\times\\
&~~~~~~~~~~\times\left(\frac{[1\, 5]^3 \langle 2\, 5\rangle  \langle 5\, n\rangle }{s_{12345} s_{2345} [1|k_{2345}k_{45}|3]}+\frac{[5\, n] \langle 2\, n\rangle ^3}{\langle 1\, 2\rangle  \langle 1\, n\rangle  \left(s_{345} [3\, n] \langle 2\, n\rangle +s_{\text{n12}} [3\, 5] \langle 2\, 5\rangle \right)}\right),\notag\\
&S_{4,D}^{+-+-}=\frac{\langle 2\, 4\rangle ^4 [n\, 5] \langle 2\, n\rangle ^3}{\langle 1\, 2\rangle  \langle 2\, 3\rangle  \langle 3\, 4\rangle  \langle 1\, n\rangle  \langle 2|k_{34}|5] \langle 2|k_{\text{n1}}k_{\text{n123}}|4\rangle \langle n|k_{1234}k_{34}|2\rangle }+\\
&~~~~~~~~~~+\frac{\langle 2\, 4\rangle ^4}{s_{234} \langle 2\, 3\rangle  \langle 3\, 4\rangle  \langle 4|k_{23}|1] \langle n|k_{1234}|5]}\left(\frac{[n\, 5] \langle n|k_{234}|1]{}^3}{s_{1234} s_{\text{n1234}} \langle 2|k_{34}k_{1234}|n\rangle }+\frac{[1\, 5]^3 \langle n\, 5\rangle }{s_{12345} \langle 2|k_{34}|5]}\right).\notag
\end{align}
In all diagrams again only one helicity configuration is non-vanishing. We tested the above result numerically against amplitude ratios and found agreement.

\acknowledgments

The author would like to thank Anastasia Volovich for suggesting this topic, initial collaboration, and useful comments regarding the draft. The author also thanks Song He for interesting discussions and encouragement on pursuing the topic during IGST16 conference. This work is supported by Simons Investigator Award \#376208 of A. Volovich.

%\acknowledgments
%
%This is the most common positions for acknowledgments. A macro is
%available to maintain the same layout and spelling of the heading.
%
%\paragraph{Note added.} This is also a good position for notes added
%after the paper has been written.
% BIBLIOGRAPHY
% use BIBTEX if you want
%\bibliographystyle{JHEP}
%\bibliography{yourBIBfiles}


%

\end{document}